\documentclass[prl,twocolumn,superscriptaddress,amsmath,amssymb,showpacs,floatfix,preprintnumbers]{revtex4-2}
\usepackage{amsmath}
\usepackage{graphicx}
\usepackage{dcolumn}
\usepackage{color}
\usepackage{physics}
\DeclareGraphicsExtensions{.png .jpg .pdf}

\begin{document}
	
	\title{Field-free-switching state diagram of perpendicular magnetization
  subjected to conventional and unconventional spin-orbit torques}
	
	\author{D. J. P. de Sousa}\email{sousa020@umn.edu}
	\affiliation{Department of Electrical and Computer Engineering, University of Minnesota, Minneapolis, Minnesota 55455, USA}
	\author{P. M. Haney}
	\affiliation{Physical Measurement Laboratory, National Institute of Standards and Technology, Gaithersburg, Maryland 20899-6202, USA}
	\author{J. P. Wang}
	\author{Tony Low}\email{tlow@umn.edu}
	\affiliation{Department of Electrical and Computer Engineering, University of Minnesota, Minneapolis, Minnesota 55455, USA}
	
	\date{ \today }
	
	\begin{abstract}
		The lack of certain crystalline symmetries in strong spin-orbit-coupled non-magnetic materials allows for the existence of uncoventional spin Hall responses, with electrically generated transverse spin currents possessing collinear flow and spin directions. The injection of such spin currents into an adjacent ferromagnetic layer can excite magnetization dynamics via unconventional spin-orbit torques, leading to deterministic switching in ferromagnets with perpendicular magnetic anisotropy. We study the interplay between conventional and unconventional spin-orbit torques on the magnetization dynamics of a perpendicular ferromagnet in the small intrinsic damping limit, and identify a rich set of dynamical regimes that includes deterministic and probabilistic switching, precessional and pinning states. Contrary to common belief, we found that there exists a critical conventional spin Hall angle, beyond which deterministic magnetization switching transitions to a precessional or pinned state. Conversely, we showed that larger unconventional spin Hall angle is generally beneficial for deterministic switching. We derive an approximate expression that qualitatively describes the state diagram boundary between the full deterministic switching and precessional states and discuss a criterion for searching symmetry-broken spin Hall materials in order to maximize switching efficiency. Our work offers a roadmap towards energy efficient spintronic devices, which might opens doors for applications in advanced in-memory computing technologies. 
		
	\end{abstract}
	
	\pacs{71.10.Pm, 73.22.-f, 73.63.-b}
	
	\maketitle
	
	\section{I. Introduction.} 
	
	Energy-efficient, external magnetic field-free switching of a thin ferromagnetic film with perpendicular magnetic anistropy (PMA) is highly desirable for novel applications in magnetic-based memories and in-memory computing technologies~\cite{ref1, ref2, ref3, ref4, refinmemory1, refinmemory2}. Modern approaches to induce field-free switching in PMA ferromagnets (FM) include spin-transfer torques (STTs) in perpendicular magnetic tunnel junctions~\cite{ref1, ref5, ref6, refpMTJ1, refpMTJ2, refpMTJ3, butler} and, more recently, spin-orbit torques (SOTs) in symmetry-broken ferromagnet/non-magnet (NM) bilayers~\cite{ref7, ref8, ref9}. 
	In typical NM/FM heterostructures, such as in W-, Pt- or Ta-based bilayers, the conventional spin Hall effect in the bulk of the NM provides an efficient mechanism to generate and inject spin currents, characterized by orthogonal spin and flow directions, into the adjacent FM.  We call the ensuing torques ``conventional spin-orbit torque'' (CSOT). For bilayers comprising NMs with a broken mirror plane symmetry, unconventional symmetry-allowed spin currents displaying a spin polarization axis aligned with the flow direction are produced~\cite{ref7, ref8, ref9, ref10}. The injection of such spin currents into an adjacent FM imparts a spin torque which we call ``unconventional spin-orbit torque" (USOT). Magnetization control through the simultaneous impact of CSOT and USOT holds promise to enable deterministic and highly energy-efficient field-free switching of PMA FMs beyond what is currently possible with regular STT or CSOT alone.
	
	We note that other improvements over CSOT-based switching have been proposed~\cite{ref11, STTSOT1, STTSOT2, STTSOT3, STTSOT4}; A recent study indicated that the presence of damping-like torques generated by symmetry breaking in a trilayer geometry can dramatically reduce the critical current density and induced deterministic switching of a PMA FM~\cite{ref11}. The enhancement in efficiency relies on the fact that such symmetry breaking damping-like torque directly competes with the intrinsic damping of the PMA FM, overcoming the damping-independent critical current density associated with damping-like CSOT switching. Hybrid STT-assisted CSOT schemes were also shown to enable efficient and deterministic switching of a PMA FM~\cite{STTSOT1, STTSOT2, STTSOT3, STTSOT4}. Here, the free PMA FM is simultaneously subjected to CSOTs, from conventional spin Hall current originating in the bulk of the NM, and symmetry breaking STTs, from the injection of out-of-plane spin polarized currents due to the presence of a fixed PMA FM layer. A common feature of these approaches is the presence of additional ferromagnetic elements to induce symmetry breaking and allow deterministic switching of the PMA FM. In this sense, CSOT-USOT-based devices have the advantage to require simpler architecture with separated write and read current paths, as it does not rely on the presence of more complicated arrangements. Additionally, it is potentially more energy-efficient due to the considerable amount of promising symmetry-broken NMs, some of which are predicted to display giant unconventional spin Hall effects~\cite{SHEDataBase}.
	
	\begin{figure*}[t]
		\includegraphics[width =0.85\linewidth]{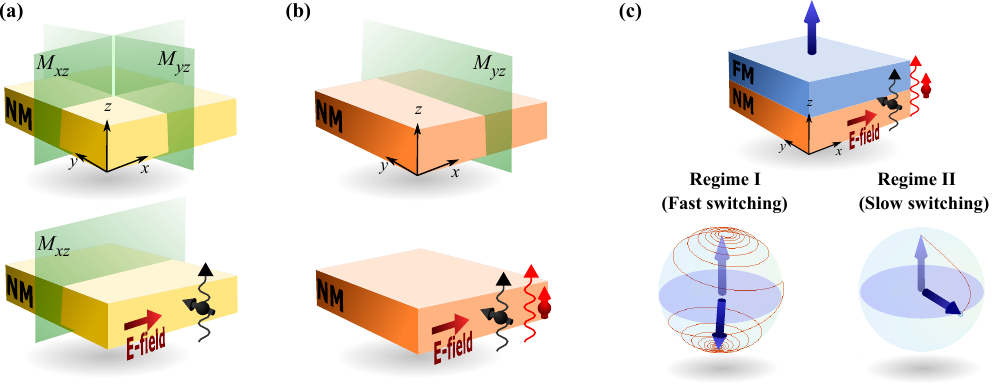}
		\caption{(Color online) Spin Hall responses of fictitious non-magnetic (NM) slabs with varying degrees of crystalline symmetries, as exemplified by the number of mirror planes, and sample magnetization trajectories of a perpendicular ferromagnet subjected to the derived spin-orbit torques. (a) NM slab with two mirror planes $M_{xz}$ and $M_{yz}$; because the mirror plane $M_{xz}$ is still a symmetry of the NM slab when the electric field is applied along the $x$ direction, the spin Hall response is constrained by the remaining symmetry to produce spin currents whose flow and spin directions are mutually orthogonal to each other and the to electric field (bottom figure).  (b) NM slab with single mirror plane $M_{yz}$; because the electric field along the $x$ direction breaks the only mirror plane available, the spin Hall response gives rise to conventional (black flowing spins) as well as unconventional (red flowing spins) spin currents, where the later is characterized by collinear spin and flow directions that are orthogonal to the electric field (bottom figure). (c) Bilayer heterostructure comprised of a ferromagnet (top layer) and a strong spin-orbit-coupled NM layer with a single crystalline $yz$ mirror plane. An electric field applied along the $x$ direction generates SOTs on the perpendicular magnetization due to conventional and unconventional spin Hall effects. The interplay between CSOT and USOT leads to the existences of distinctive dynamical regimes. The bottom figures illustrate dynamical regimes where full switching and full in-plane damping of the magnetization are expected. We note that switching to the $-\hat{\textbf{z}}$ direction in regime I is faster than switching in regime II in the small intrinsic damping limit.}
		\label{fig1}
	\end{figure*}
	
	    Theoretical and experimental developments have shown that field-like SOTs can directly impact the dynamical state of a magnetization in a bilayer geometry~\cite{ref12, ref13, ref14, ref15, ref16}. For instance, field-like CSOTs have been shown to greatly influence the threshold switching condition by introducing a half-integer power dependence of the damping constant to the critical switching current~\cite{ref12}.	Recent experiments revealed the existence of large field-like CSOTs in Ru$_2$Sn$_3$/CoFeB bilayers~\cite{ref13}, which was also shown to substantially modify the dynamical state of the in-plane FM layer. From these developments, one might anticipate that field-like CSOTs might also play an important role in determining the dynamical states of a PMA FM simultaneously subjected to CSOT and USOTs. Hence, the evaluation of the field-free switching state diagram of a PMA FM, including the impact of field-like CSOTs and USOTs, becomes crucial in accessing the potential of such symmetry-broken bilayers for highly energy-efficient switching applications, which has not yet been investigated in detail. 
	
	In this work, we numerically compute field-free state diagrams for a perpendicular magnetization simultaneously subjected to field-like and damping-like CSOT and USOT. Our analysis reveals unanticipated dynamical regimes characterized by stable precessional and pinned states whose existences are strongly connected to the sign and strength of the field-like CSOT. We find that energy-efficient field-free switching might not be achieved by simultaneously maximizing the spin Hall angles associated with CSOTs and USOTs. Instead, while large spin Hall angles associated with USOT are always beneficial, there is an ideal CSOT spin Hall angle for which highly energy-efficient switching takes place. Our work offers a systematic pathway to realizing energy-efficient switching of a perpendicular magnetization in symmetry-broken bilayer heterostructures.  
	
	\section{II. Theoretical model} 
			\begin{figure}[t]
		\includegraphics[width =0.9\linewidth]{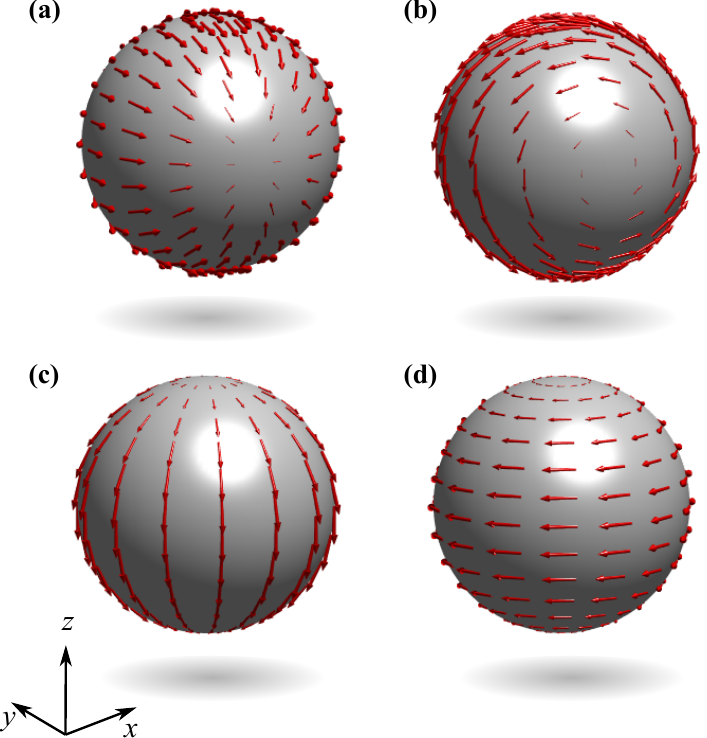}
		\caption{(Color online) Depiction of the spin-orbit field-like and damping-like torques over the unit sphere. The conventional damping-like and field-like torques are shown panels (a) and (b), respectively. The unconventional damping-like and field-like spin-orbit torques are depicted in panels (c) and (d), respectively. These correspond to the four terms in the spin-orbit torque density in the Landau-Lifshitz-Gilbert equation, as defined in Eq.~(\ref{eq1}).}
		\label{fig0}
	\end{figure}
	We study a bilayer heterostructure comprised of a thin PMA FM overlaid on a strongly spin-orbit coupled symmetry-broken NM. The reduced symmetry is assumed to be an intrinsic property of the crystal structure of the NM. The importance of such a feature is illustrated in Figs.~\ref{fig1}(a) and (b), where we compare the spin Hall response of two fictitious NMs with distinct degrees of symmetry; For the NM slab with two mirror planes, $M_{xz}$ and $M_{yz}$ [shown in Fig.~\ref{fig1}(a)], only $M_{xz}$ remains a symmetry of the system under an applied electric field along the $x$ direction. In this scenario, the presence of $M_{xz}$ constrains the spin Hall response to produce spin currents with mutually orthogonal spin and flow directions (black flowing spins). On the other hand, the symmetry broken NM slab with a single mirror plane, $M_{yz}$ [shown in Fig.~\ref{fig1}(b)], is left without any mirror symmetry under the action of an electric field applied along the $x$ direction. Consequently, the spin Hall response is not constrained, which enables unconventional spin Hall currents with collinear spin and flow directions (red flowing spins) in addition to the conventional response. 
	
	A symmetry-broken bilayer heterostructure composed of a PMA FM and the NM with a single mirror plane is illustrated in Fig.~\ref{fig1}(c). Here, an in-plane electric field generates a SOT on the magnetization of the FM due to a flow of spin-polarized electrons originating from the spin Hall effect in the bulk of the NM. Assuming an electric field applied along the $x$ direction, the generated spin currents in the bulk of the NM are $Q_z^y = \sigma_{zx}^y E_x$ and $Q_z^z = \sigma_{zx}^z E_x$, where the subscripts refer to the flow direction and the superscripts to the spin-quantization axis, as sketched in Fig.~\ref{fig1}(c). The former component, associated with $\sigma_{zx}^y$, is the conventional spin Hall response, where field, spin and flow directions are mutually orthogonal to each other. The latter component, associated with $\sigma_{zx}^z$, is the unconventional response that only exists due to the lack of a $xz$ mirror plane in the NM. The transfer of spin angular momentum from the spin Hall currents to the magnetization imparts a total SOT, which can be classified as damping- and field-like CSOT and USOT. A promising example of such a system is that of 1T'-WTe$_2$/Py bilayers, whose USOT was recently observed in experiments~\cite{ref7, ref9, ref17}. Symmetry-dependent USOT-based switching was also observed in CuPt/CoPt bilayers~\cite{CuPtCoPt}, which is equally well described by the model presented here inspite of the richer lattice symmetries of L1$_{1}$-ordered Cu(Co)Pt crystals. In our study, the perpendicular anisotropy axis of the FM layer coincides with the z axis. 
	
	The total SOT torque density is
	
	%We study a bilayer heterostructure comprised of a thin FM layer with perpendicular magnetic anisotropy and a NM layer with broken in-plane symmetry and strong spin-orbit coupling. The bilayer heterostructure has lowered symmetry due to the presence of the NM, whose crystal structure lacks a $xz$ mirror plane as shown in Fig.~\ref{fig1}(a). A promising example is that of 1T'-WTe$_2$/Py bilayers, whose USOT was recently observed in experiments~\cite{ref7, ref9, ref17}. Here, the perpendicular anisotropy axis of the FM layer coincides with the z axis. In this system, an in-plane electric field generates a SOT on the magnetization of the FM due to a flow of spin-polarized electrons originating from the spin Hall effect in the bulk of the NM. Assuming an electric field applied along the $x$ direction, the generated spin-currents in the bulk of the NM are $Q_z^y = \sigma_{zx}^y E_x$ and $Q_z^z = \sigma_{zx}^z E_x$, where the subscripts refer to the flow direction and the superscripts to the spin-quantization axis, as sketched in Fig.~\ref{fig1}(b). The former component, associated with $\sigma_{zx}^y$, is the conventional spin Hall response, where field, spin and flow directions are mutually orthogonal to each other. The latter component, associated with $\sigma_{zx}^z$, is the unconventional response that only exists due to the lack of a $xz$ mirror plane. The electrically generated spin-currents impart a SOT on the magnetization of the FM layer whose torque density is
	\begin{eqnarray}
		\boldsymbol{\mathcal{N}} = \tau_{\rm{DL}}^y \hat{\textbf{m}} \times (\hat{\textbf{m}} \times \hat{\textbf{y}}) + \tau_{\rm{FL}}^y \hat{\textbf{m}} \times \hat{\textbf{y}} \nonumber \\
		+ \tau_{\rm{DL}}^z \hat{\textbf{m}} \times (\hat{\textbf{m}} \times \hat{\textbf{z}}) + \tau_{\rm{FL}}^z \hat{\textbf{m}} \times \hat{\textbf{z}},
		\label{eq1}
	\end{eqnarray}
	where $\tau_{\rm{DL}}^{y(z)} = (\hbar/2e)\theta_{y(z)} J$ and $\tau_{\rm{FL}}^{y(z)} = \beta \tau_{\rm{DL}}^{y(z)}$ are the
	damping-like and field-like contributions, respectively,
	with $J$ being the applied current density. For simplicity we assume that $\beta$ is the same for both $y$ and $z$ spin components. The conventional (unconventional) spin Hall angle is defined as $\theta_{y(z)} = (2e/\hbar) \sigma_{zx}^{y(z)}/\sigma_{xx}$. The damping- and field-like CSOT and USOT are depicted over the unit sphere in Fig.~\ref{fig0}; The red arrows represent the specific normalized SOT acting on the magnetization whose direction is defined by a single point on the surface of the sphere. While the damping-like CSOT (USOT) relaxes the magnetization towards the $-\hat{\textbf{y}}$ ($-\hat{\textbf{z}}$) direction, the associated field-like torque causes the magnetization to undergo precession around the $\hat{\textbf{y}}$ ($\hat{\textbf{z}}$) direction.
	
	The Landau-Lifshitz-Gilbert (LLG) equation governing the magnetization dynamics reads
	\begin{eqnarray}
		\frac{d\hat{\textbf{m}}}{dt} = -\gamma \hat{\textbf{m}} \times \textbf{H}_{\rm{eff}} + \alpha \hat{\textbf{m}} \times \frac{d\hat{\textbf{m}}}{dt} + \frac{\gamma}{\mu_0 M_{\rm{S}}d} \boldsymbol{\mathcal{N}},
		\label{eq2}
	\end{eqnarray}
	where $\hat{\textbf{m}} = \textbf{M}/M_{\rm{S}}$ with $\textbf{M}$ being the magnetization of the FM film with saturation value $M_{\rm{S}}$, $\gamma$ is the gyromagnetic ratio, $\alpha$ is the intrinsic damping parameter, $\mu_0$ is the vacuum permeability and $d$ is the thickness of the ferromagnetic film. The perpendicular magnetic anisotropy is described through the effective field $\textbf{H}_{\rm{eff}} = (2K_{\rm{eff}}/\mu_0 M_{\rm{S}})(\hat{\textbf{m}} \cdot \hat{\textbf{z}})\hat{\textbf{z}}$ with $\hat{\textbf{z}}$ being perpendicular to the interface plane. The effective anisotropy parameter $K_{\rm{eff}}$ is related the interfacial anisotropy constant $K_{\rm{i}}$ as $K_{\rm{eff}} = K_{\rm{i}}/d - \mu_0M_{\rm{S}}^2 /2$. In the following, we numerically solve Eq. (2) for the time evolution of $\hat{\textbf{m}}$ and explore the simultaneous impact of $\theta_{y}$ and $\theta_{z}$ on the magnetization dynamics.
	
	\section{III. Preliminary discussion}
	
	In general, the qualitative aspects of the magnetization dynamics expected for such system depends upon the relative magnitude of the various parameters involved.  We start by distinguishing two main dynamical regimes, which will be referred to as regimes I and II [in the bottom panel of Fig.~\ref{fig1}(c)]. In regime I, the dynamics is characterized by a switching event similar to that induced by spin-transfer torque in perpendicular magnetic tunnel junctions~\cite{refpMTJ2}, i.e. the magnetization relaxes to a stable focus at $\hat{\textbf{m}} \approx -\hat{\textbf{z}}$ as shown in Fig.~\ref{fig1}(c). Here, the damping-like torque attributed to $\theta_z$ dominates the dynamics by overcoming the intrinsic damping of the FM layer, whereas the overall effect of $\theta_y$ is to modify the precessional rates during the dynamical evolution and to cause a small tilt of the final magnetization configuration away from $-\hat{\textbf{z}}$. The latter feature and intrinsic damping ensures deterministic switching to the $\hat{\textbf{m}} = -\hat{\textbf{z}}$ configuration when the applied current pulse vanishes. 
	
	%{\it We emphasize that the existence of the dynamical regime I is not confined to the situation where $\theta_y \ll \theta_z$, as one might naively conclude from the above description. In fact, our analysis indicates that dynamics with regime I characteristics also takes place when $\theta_y > \theta_z$ in some situations. This is due to the fact that while $\tau_{\textrm{DL}}^y$ competes with the perpendicular anistropy, $\tau_{\textrm{DL}}^z$ has to overcome the much smaller intrinsic damping of the ferromagnetic layer. This observation ensures that the definition of regime I is not confined to a limiting case, as will be seen later.}
	
	In regime II, the dynamics is characterized by a steady state where the magnetization is damped towards $\hat{\textbf{m}} \approx -\hat{\textbf{y}}$, as shown in Fig.~\ref{fig1}(d). In this regime, the damping-like torque attributed to $\theta_y$ dominates the dynamics, whereas the $\theta_z$ only causes a small tilt of the final magnetization configuration towards the $-\hat{\textbf{z}}$ direction. Here, deterministic switching to the $\hat{\textbf{m}} = -\hat{\textbf{z}}$ configuration might not occur when the applied current passing through the NM is turned off, particularly if thermal fluctuations are strong enough. Hence, while regime I supports deterministic switching of the perpendicular magnetization, regime II does not necessarily. If thermal fluctuations are weak, regime II may also lead to switching after the current pulse has vanished. Here, the intrinsic damping of the FM layer guarantees that the perpendicular magnetization relaxes from $\approx -\hat{\textbf{y}}$ towards $-\hat{\textbf{z}}$ if there exists a small downward tilting due to $\theta_z$. However, this process is very slow due to the smallness of intrinsic damping parameters in typical transition metal-based FM layers. Therefore, switching in regime I is much faster than switching in regime II in the small damping limit.

	The distinction between regimes I and II provide a starting point for understanding the system behavior, however we will see that the full spectrum of system behavior includes several other regimes, such as precessional and pinned states.  One key factor determining the system behavior is the relative orientation between the magnetization and the spin direction of the incoming spin current. For incoming spin oriented perpendicular to the magnetization, the damping-like torque competes directly with the torque from anisotropy. On the other hand, for nearly collinear magnetization and incoming spin, the damping-like torque competes with the torque from magnetic damping, which is weaker by a factor of $\alpha$.  Accordingly, the critical current for switching is a factor $\alpha$ smaller for collinear incoming spin and magnetization, compared to perpendicular incoming spin and magnetization.
	
	We find that the field-like CSOT plays a key role in the dynamics.  This is due to the fact that the field-like CSOT tilts the magnetization away from the anisotropy axis towards the $y$-axis by a small amount $\theta_{\textrm{tilt}}$ (see Fig.~\ref{figPaul}). The direction of the tilt depends on the sign of $\beta$.   Figure~\ref{figPaul} shows different scenarios.  If the magnetization is tilted toward (away) from the incoming spin, the critical current decreases (increases), due to the factors described in the previous paragraph.  For this reason, we will find that the field-like torque plays an important role in determining the magnetic dynamics.

	\section{IV. Current-induced dynamical state transitions}

	\begin{figure}[t]
		\includegraphics[width =\linewidth]{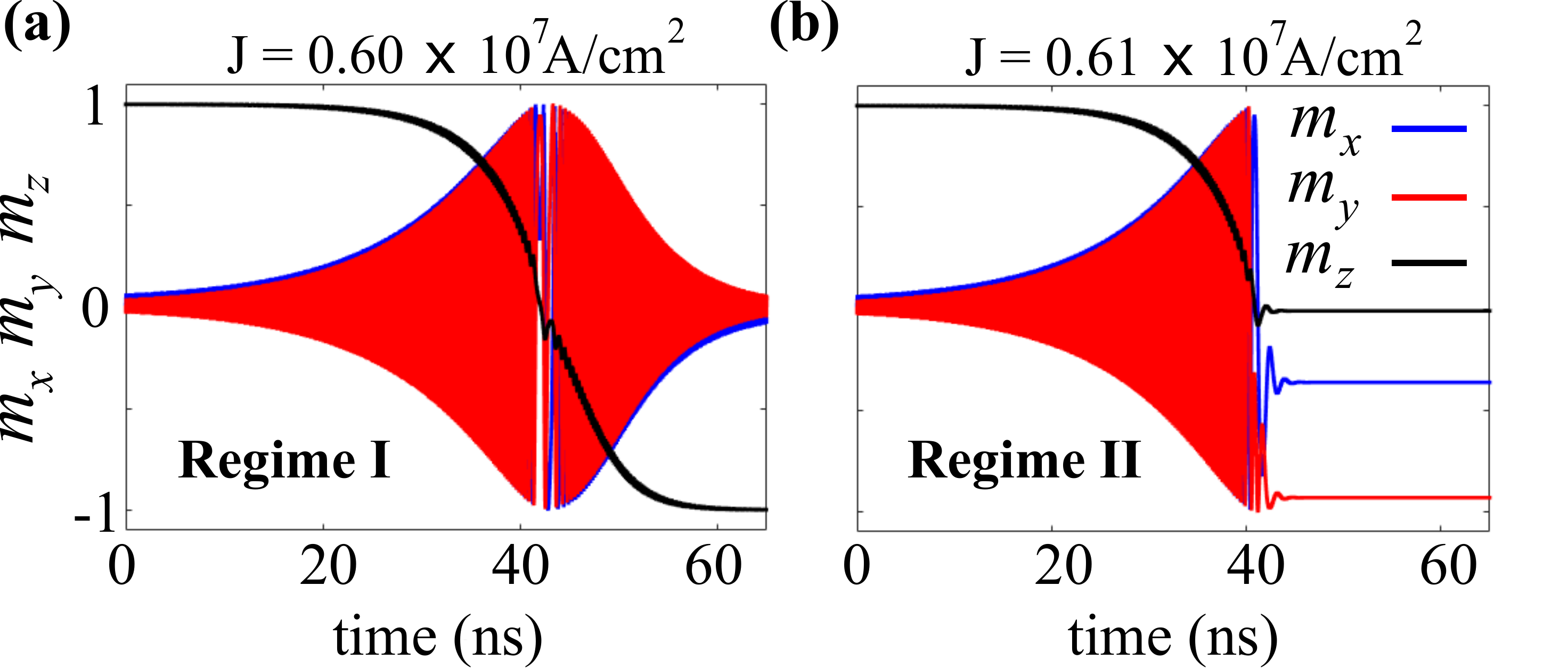}
		\caption{(Color online) Time evolution of the components of the magnetization at a \textit{negative} field-like to damping-like torque ratio of $\beta = -0.2$. Upon increasing the applied current density from (a) $J = 0.60 \times 10^7$ A/cm$^2$ to (b) $J = 0.61 \times 10^7$ A/cm$^2$, with $\theta_y = 0.6$ and $\theta_z = 0.05$, a transition from dynamical regime I to regime II takes place.}
		\label{fig2}
	\end{figure}
	
	In this section we show that a current-induced transition between regimes I and II takes place and is highly affected by the strength and sign of the field-like to damping-like torque ratio as quantified by $\beta$~\cite{footnote}. We focus on describing magnetization dynamics of PMA FMs with small intrinsic damping, $\alpha = 0.001$. We assume the following parameters throughout: $d = 1$ nm, $K_{\rm{eff}} = 2 \times 10^5$ J/m$^3$ and $M_{\rm{S}} = 1.27\times 10^6$ A/m. 
	
	First, we study the current-induced transition at \textit{negative} $\beta$; we take $\beta = -0.2$. Figure~\ref{fig2} displays the time evolution of the magnetization components due to CSOT and USOT quantified by the spin Hall angles $\theta_y = 0.6$ and $\theta_z = 0.05$, respectively, at two current densities. At lower currents ($J = 0.60 \times 10^7$ A/cm$^2$), magnetization reversal takes place with $\hat{\textbf{m}}(t = +\infty) \approx -\hat{\textbf{z}}$, with initial condition $\hat{\textbf{m}}(t = 0) \approx \hat{\textbf{z}}$. This situation is shown in Fig.~\ref{fig2}(a) and characterizes a magnetization dynamics of regime I. Upon increasing the applied current density, the switched steady state slowly deviates from the $\approx -\hat{\textbf{z}}$ until an abrupt transition to a new in-plane steady state with $\hat{\textbf{m}}(t = +\infty) \approx -\hat{\textbf{y}}$ takes place at a critical current density. The time evolution of the magnetization components for a slightly larger current density of $J = 0.61 \times 10^7$ A/cm$^2$ (the critical current) is shown in Fig.~\ref{fig2}(b). Therefore, we conclude that a \textit{direct} transition between regimes I and II can be induced by current in the negative $\beta$ case. The evolution of the transition point with the strength of $\beta$ will be addressed in upcoming sections.
	
		\begin{figure}[t]
		\includegraphics[width =\linewidth]{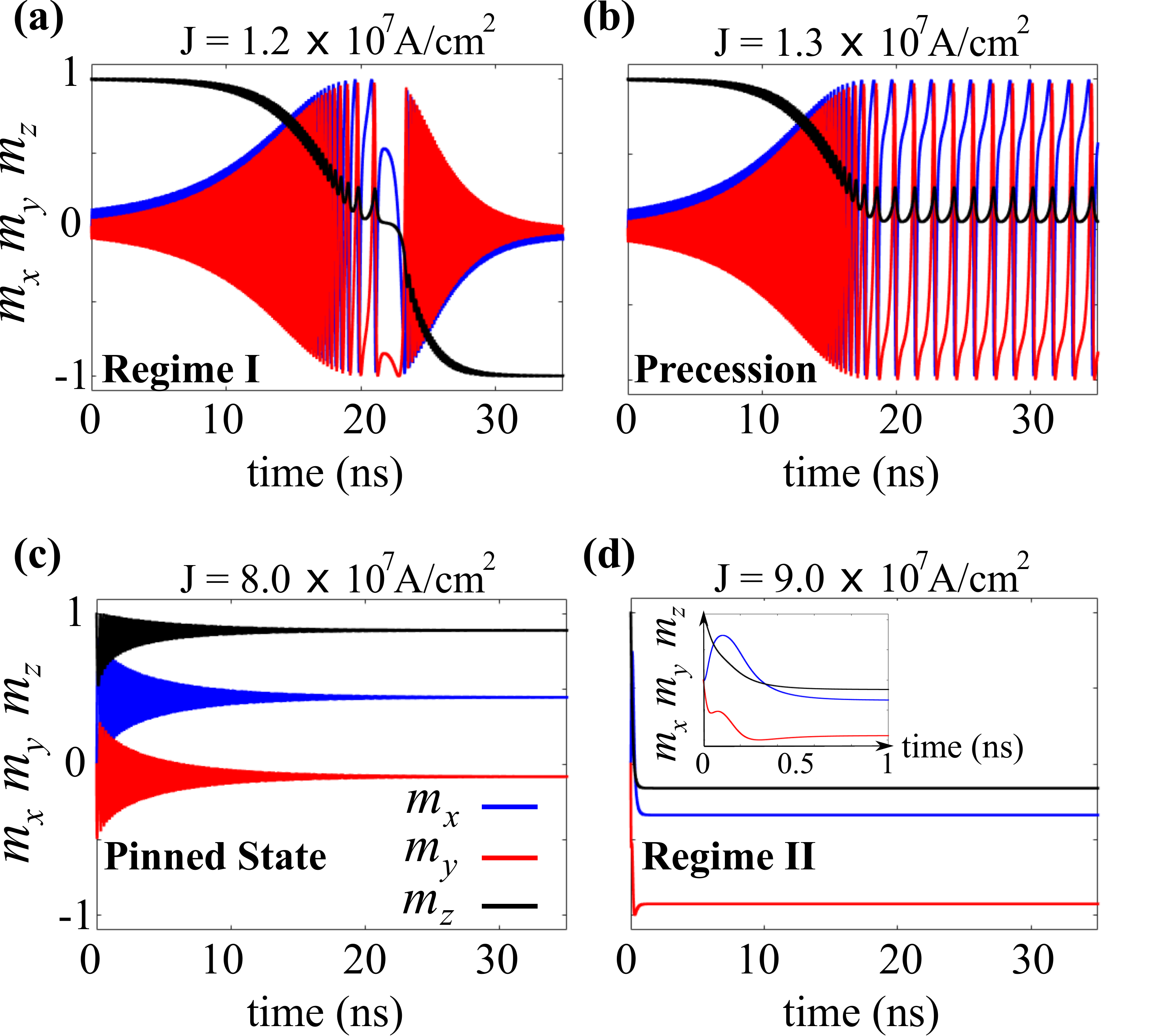}
		\caption{(Color online)  Time evolution of the components of the magnetization at a \textit{positive} field-like to damping-like torque ratio of $\beta = +0.2$. Here, transitions between dynamical regimes I and II take place through intermediate stable precessional and pinned state regimes: (a) $J = 1.2 \times 10^7$ A/cm$^2$ (b) $J = 1.3 \times 10^7$ A/cm$^2$, (c) $J = 8.0 \times 10^7$ A/cm$^2$ and (d) $J = 9.0 \times 10^7$ A/cm$^2$, with $\theta_y = 0.6$ and $\theta_z = 0.05$.}
		\label{fig3}
	\end{figure} 
	
	We now focus on the current-induced dynamical regime transition considering \textit{positive} field-like to damping-like torque ratios. For the sake of comparison, all parameters are assumed to be the same as the previous calculation except for the sign of $\beta$, i.e., $\beta = +0.2$. Figure~\ref{fig3} shows the time evolution of the magnetization due to CSOT and USOT at increasing current densities. We observed in Fig.~\ref{fig3}(a) that the dynamics excited by a current density of $J = 1.2 \times 10^7$ A/cm$^2$ falls within regime I. This is in sharp contrast to the $\beta < 0$ case, where the dynamics induced by the same current density resides in regime II. 
	
	An \textit{indirect} current-induced transition to regime II that takes place at $\beta > 0$, in contrast with the $\beta < 0$ scenario. Here, a larger current-density induces transitions to intermediate dynamical regimes before the full transition to regime II. This is shown in Figs.~\ref{fig3}(b) and (c), where transitions to a stable precessional and steady states occur upon increasing the current density to $J = 1.3 \times 10^7$ A/cm$^2$ and subsequently to $J = 8.0 \times 10^7$ A/cm$^2$, respectively. Transition to regime II occurs at much larger current-densities, as shown in Fig.~\ref{fig3}(d) for $J = 9.0 \times 10^7$ A/cm$^2$. The inset shows in detail the typically damped time evolution of the magnetization components in the $\beta > 0$~\cite{ref12}, which is qualitatively distinct from the regime II oscillatory behavior for the $\beta < 0$ in Fig.~\ref{fig2}(b). 
	
	The above discussion indicates that the sign of $\beta$ is crucial in determining the condition for full reversal of a perpendicular magnetization simultaneously subjected to CSOT and USOT. The observation that larger current-densities might actually be detrimental to an already switched configuration by inducing transitions to a precessional state shows that a full state diagram covering the various dynamical regimes over a given parameter space is necessary for further guiding experimental work in this field. In the following, we construct and discuss the main characteristics of relevant state diagrams.
	
	\section{V. Dynamical state diagrams}
	
	We provide insights into the various dynamical regimes of a perpendicular magnetization subjected to CSOT and USOT by constructing two main state diagrams: A current-density versus $\beta$ at fixed conventional and unconventional spin Hall angles ($\theta_y = 0.6$ and $\theta_z = 0.05$) and a $\theta_y$ vs $\theta_z$ state diagram at fixed $J = 3.0 \times 10^7$ A/cm$^2$ and $\beta = +0.2$. The diagrams are constructed by numerically obtaining the boundaries between distinct dynamical regimes over the given parameter space. The boundaries are easily identified through sharp changes in magnetization dynamics at critical parameter values, as shown in panels (a) and (b) in Figs.~\ref{fig2} and \ref{fig3}. 
	
\begin{figure}[t]
		\includegraphics[width =\linewidth]{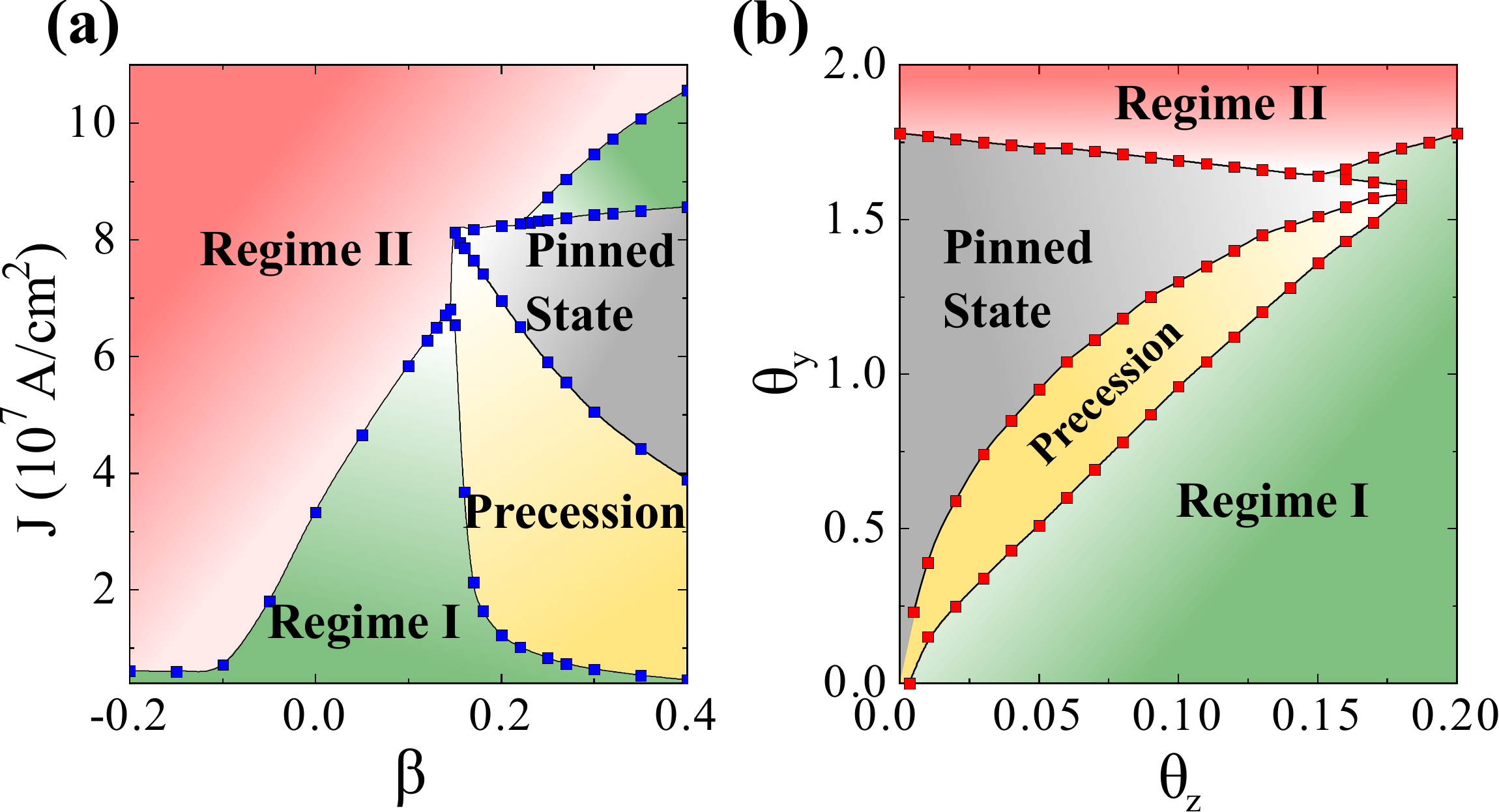}
		\caption{(Color online)  Switching diagrams of a perpendicular magnetization subjected to both CSOT and USOT. (a) current density $\times$ field-like to damping-like torque ratio switching diagram at fixed $\theta_y = 0.6$ and $\theta_z = 0.05$. (b) $\theta_y$ vs $\theta_z$ switching diagram at fixed $J = 3 \times 10^7$ A/cm$^2$ and $\beta = +0.2$. Distinct dynamical regimes are represented by different colors.  The boundary lines were obtained numerically by inspecting the full magnetization dynamics using the LLG equation. }
		\label{fig4}
	\end{figure}

\begin{figure}[t]
		\includegraphics[width =\linewidth]{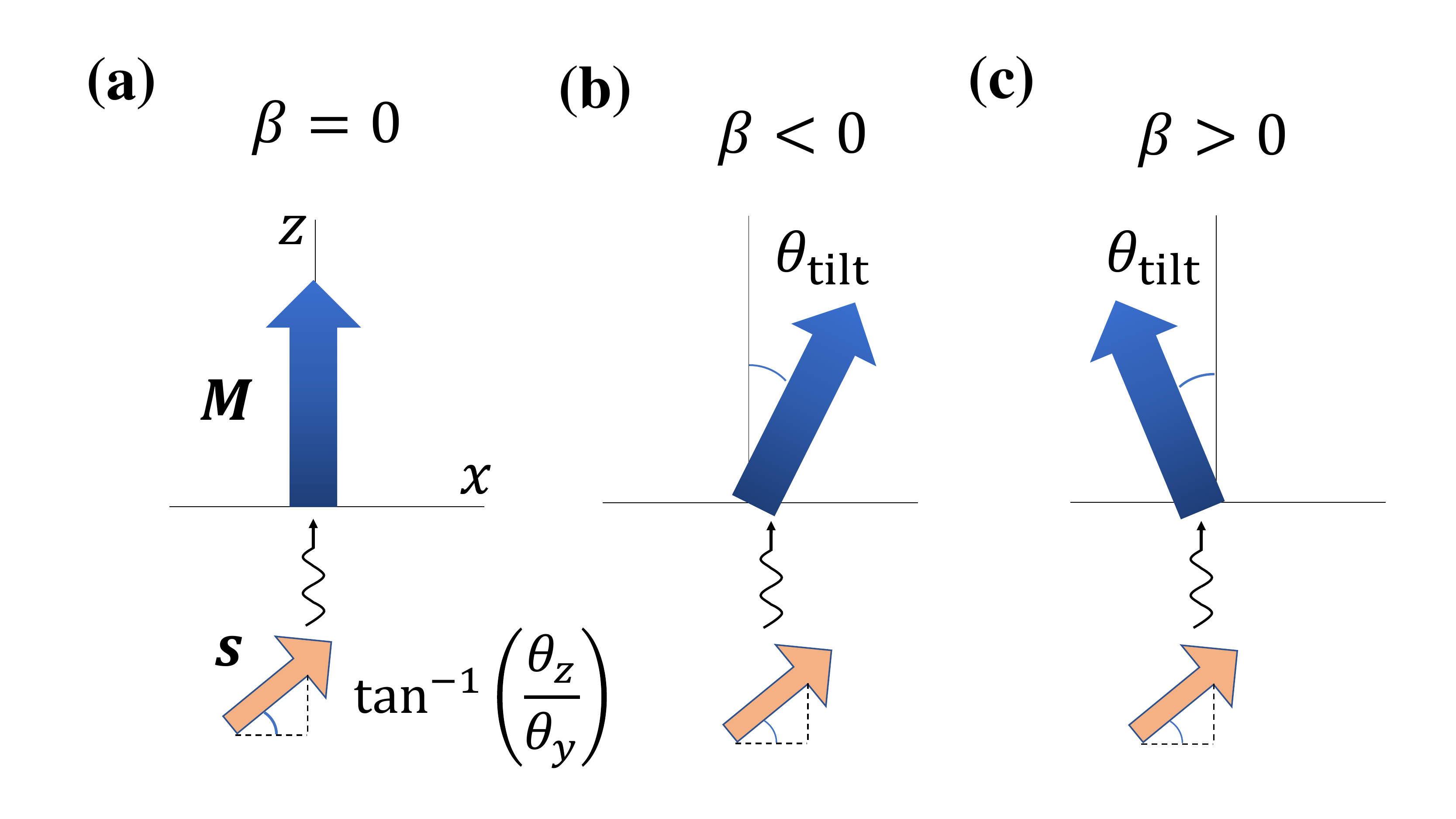}
		\caption{(Color online)  A schematic of the key factors controlling the magnetic dynamics. The magnetization (blue arrow) has an effective easy-axis which is is determined by the magnetic anisotropy and the field-like torque, which tilts the magnetization from the z-axis by an amount $\theta_{\textrm{tilt}}$. The incoming spin direction (orange arrow) is determined by $\theta_z/\theta_y$.  The squiggly arrow denotes velocity direction of spin current.  The critical switching current is decreased when the magnetization and incoming spin are more aligned, as in (b).}
		\label{figPaul}
	\end{figure}

	The state diagrams are presented in Fig.~\ref{fig4}, where distinct dynamical regimes are highlighted by different colors. The symbols demarcating the boundaries are the numerically obtained critical transition points. The $J$ vs $\beta$ state diagram in Fig.~\ref{fig4}(a) reveals a single transition boundary between regimes I and II at all negative $\beta$ cases we considered. Here, the critical current-density necessary to induce a dynamical state transition to regime II decreases with increasing $|\beta|$, consistent with previous predictions~\cite{ref12}. At $\beta > 0$, there also exists a single transition boundary between regimes I and II for $\beta < 0.145$, but with the critical current-density being an increasing function of $|\beta|$. Beyond the critical value $\beta \approx 0.145$ there appears intermediate dynamical regimes characterized by stable precessional and steady states. Hence, we conclude that there exists a critical $\beta > 0$ separating direct and indirect transitions between regimes I and II, beyond which our results indicate an expansion (contraction) of the steady state (precessional state) region with further increase $\beta$. We note that the large current-density window within which the precessional states exist at a given $\beta > 0.145$ might be ideal for applications in nano-oscillators that are robust to current-density variations. We stress that further studies are necessary to address the potential of the precessional states for applications, which is beyond the scope of this work. 
	
	In Fig.~\ref{fig4}(b) we show the spin Hall angle state diagram at fixed $J = 3 \times 10^7$ A/cm$^2$ and $\beta = 0.2$. Here, the $\beta$ value has been chosen to enable the analysis of the evolution of all distinct dynamical regimes with $\theta_y$ and $\theta_z$. The state diagram shows that while the extent of regime II is limited to large $\theta_y$ values ($\theta_y > 1.5$), regime I-type of dynamics is possible for unconventional spin Hall angles as small as $\theta_z \approx 0.01$ at sufficiently small $\theta_y$. Notice that regime I is only accessible at finite $\theta_z$. Our numerical results indicate that spin Hall angle ratios of $\theta_z/\theta_y \approx 10 \%$ are enough to induced a regime I-type of dynamics, i.e., full magnetization reversal of a perpendicular magnetization. The possibility of full magnetization switching at small $\theta_z/\theta_y$ ratios is tightly connected to fact that, while CSOT are required to overcome the anistropy field for full switching, the damping-like USOT has to counteract the much smaller intrinsic damping. In this sense, the USOT is promoted to the role of being the main switching agent, whereas the conventional one provides an assisting mechanism. The above picture is supported by experimentally observed magnetization switching in 1T'-WTe$_2$/Py bilayers where $\theta_z/\theta_y \approx 24~ \%$~\cite{ref17}. Other, symmetry-broken NMs potentially suitable for observing the effects mentioned above are BiTe$_3$, Ni$_2$P$_6$W$_4$ and Ba$_2$C$_4$S$_4$N$_4$, to cite a few, all of which display giant $\sigma_{zx}^{z}$ above 1000 ($\hbar/2e$)$\Omega^{-1}$cm$^{-1}$~\cite{SHEDataBase}. 
	
	Figure~\ref{fig4}(b) also indicates an unanticipated scenario in which large conventional spin Hall angles become detrimental to deterministic switching of a perpendicular magnetization; Upon further increasing $\theta_y$ beyond the critical boundary line at fixed $\theta_z$, full magnetization reversal is suppressed due to transitions to precessional and steady states regimes. Hence, we obtain the unexpected result that larger conventional spin Hall angles might be detrimental to switching in symmetry-reduced bilayer heterostructures.
	
	In the following section, we provide a qualitative description of the above behavior from an analytical toy model perspective. Our analysis indicates that upon increasing $\theta_y$, an effective easy-axis tilting induced by the field-like CSOT develops, altering the switching threshold condition. The tilt depends on the sign of $\beta$, providing an explanation for the disparate behavior shown in Fig.~\ref{fig4} (a) for $\beta > 0$ and $\beta < 0$.
	
		\section{VI. Impact of field-like torques: A model description}

	We next present an analytical model to describe some of the qualitative features of the numerical data presented in the previous section.  We begin by noting that the tilt of the magnetization is given by:
	\begin{eqnarray}
		\tan \theta_{\rm{tilt}} = \beta \frac{\hbar}{4e} \theta_y \frac{J}{K_{\rm{eff}}d}.
		\label{eq}
	\end{eqnarray}
	
	We assume that the only effect of the field-like torque is to tilt the uniaxial anisotropy axis by a small amount of $\theta_{\rm{tilt}}$\cite{footnote2}. Within this approximation, the critical switching current due to CSOT and USOT is obtained by rotating the coordinate system by $\theta_{\rm{tilt}}$ so as to align the new $z$ axis with the tilted main anisotropy axis. The net effect on the spin quantization axis of the incoming spin current amounts to taking the effective spin hall angles $\tilde{\theta}_z$ and $\tilde{\theta}_y$ according
	\begin{eqnarray}
		\left(
		\begin{array}{c}
			\tilde{\theta}_z \\
			\tilde{\theta}_y 
		\end{array}\right)  
		= 
		\left(
		\begin{array}{cc}
			\cos\theta_{\rm{tilt}} & -\sin\theta_{\rm{tilt}} \\
			\sin\theta_{\rm{tilt}} & \cos\theta_{\rm{tilt}} 
		\end{array}\right)
		\left(
		\begin{array}{c}
			\theta_z \\
			\theta_y 
		\end{array}\right),
		\label{eq3}
	\end{eqnarray} 
	where the sign of $\theta_{\rm{tilt}}$ is determined by the sign of the field-like to damping-like torque ratio $\beta$. Hence, for $\beta < 0$ we take $\theta_{\rm{tilt}} \rightarrow - \theta_{\rm{tilt}}$ in Eq.~(\ref{eq3}).
	
	The critical switching current in the rotated coordinate systems is approximately given by~\cite{ref11}
	\begin{eqnarray}
		J_{\rm{sw}}(\theta_{\rm{tilt}}) = \displaystyle \frac{4e K_{\rm{eff}}d}{9\alpha \hbar}A(\theta_{\rm{tilt}}) B(\theta_{\rm{tilt}}) \frac{\tilde{\theta}_z}{\tilde{\theta}_y^2},
		\label{eq4}
	\end{eqnarray}
	where we have defined
	\begin{subequations}
		\begin{eqnarray}
			A(\theta_{\rm{tilt}}) = \displaystyle -1 + \sqrt{1 + 6\alpha^2 \left(\frac{\tilde{\theta}_y}{\tilde{\theta}_z}\right)^2},
			\label{5a}
		\end{eqnarray}
		\begin{eqnarray}
			B(\theta_{\rm{tilt}}) = \sqrt{3 + \frac{12}{2 + A(\theta_{\rm{tilt}})}}.
			\label{5b}
		\end{eqnarray}
	\end{subequations}
	
	The form for the critical current can be intuitively understood in some limiting cases that we describe next. These limiting cases provide a way to qualitatively understand the impact of field-like torques on the dynamical state of the perpendicular magnetization. 
	
	We first consider $\beta > 0$ and assume that the incoming spin is perpendicular to the tilted anisotropy axis.   The effective spin Hall angles in the rotated coordinates system become $\tilde{\theta}_y \rightarrow \theta_y /\cos(\theta_{\rm{tilt}})$ and $\tilde{\theta}_z \rightarrow 0$. Hence, the excited magnetization dynamics in the rotated coordinates system is dominated by a damping-like torque due to incoming spin currents with polarization direction transverse to the main tilted anisotropy axis. This scenario is similar to that in bilayer heterostructures under the action of CSOTs, which exhibit critical switching currents that are independent of $\alpha$~\cite{ref11, ref12, ref18}. Accordingly, we find the critical switching current is given by:
	\begin{eqnarray}
		J_{\rm{sw}}^{\beta > 0} \approx \displaystyle \frac{4\sqrt{2}e}{3 \hbar}\frac{K_{\rm{eff}}d}{\theta_y},
		\label{eq6}
	\end{eqnarray}
	which is larger than $J_{\rm{sw}}(\theta_{\rm{tilt}} = 0)$, in Eq.~(\ref{eq4}), by a factor of $\approx \alpha^{-1}$.  
	
	We next consider $\beta < 0$ and assume the incoming spin direction is collinear to the tilted anisotropy axis.  The effective spin Hall angles in the rotated coordinates system become $\tilde{\theta}_y \rightarrow 0$ and $\tilde{\theta}_z \rightarrow \theta_y /\cos(\theta_{\rm{tilt}})$.  Here, the magnetization dynamics is dominated by damping-like torques due to incoming spin currents with polarization direction aligned with the main tilted anisotropy axis. In this situation, the damping-like torque directly competes with intrinsic damping, resulting in critical switching current proportionals to $\alpha$.  Indeed, the critical switching current is 
	\begin{eqnarray}
		J_{\rm{sw}}^{\beta < 0}  \approx \displaystyle \frac{4e}{\hbar}\alpha \frac{K_{\rm{eff}}d}{\theta_z},
		\label{eq7}
	\end{eqnarray}
	which is smaller than $J_{\rm{sw}}(\theta_{\rm{tilt}} = 0)$ of Eq.~(\ref{eq4}) in the small damping limit.  
	
	%We can intuitively understand the result summarized in Eq.~($\ref{eq6}$) and Eq.~($\ref{eq7}$) as follows: In the $\beta > 0$ limit, $\tan(\theta_{\rm{tilt}}) \rightarrow \theta_z/\theta_y$,. In the $\beta < 0$ limit, $\tan(\theta_{\rm{tilt}}) \rightarrow \theta_y/\theta_z$,  
	
	The above discussion indicates that the sign and strength of the conventional field-like to damping-like torque ratio is crucial in determining the current threshold condition for switching. This observation results in the following picture; The amount of anisotropy axis tilting due to the conventional field-like torque is determined by $\beta$ as well as the applied current density $J$, as described in Eq.~(\ref{eq}). If $\beta > 0$ ($\beta < 0$), a new switching current threshold will be established by the induced tilting, $\theta_{\textrm{tilt}}$.
	%, with the approximate value given in Eq.~($\ref{eq6}$) (Eq.~($\ref{eq7}$)) in the limit $\tan(\theta_{\rm{tilt}}) \rightarrow \theta_z/\theta_y$ ($ \rightarrow \theta_y/\theta_z$). 
	Hence, the applied current density $J$ determines a new current switching threshold $J_{\textrm{sw}}^{\beta > 0 (\beta < 0)}$ in accordance to the $\beta$.
	% This is only possible due to the existence of finite field-like CSOTs. 

	As a consequence, the threshold condition to excite magnetization dynamics within regime I will depend on $\theta_z$ and $\theta_y$, as well as on $J$ and $\beta$. This implies that an initially stable regime I type-of-dynamics induced by a current density $J$ might not be made possible by further increasing the applied current density due to a larger switching threshold induced by the tilting.   This is also possible for systems with larger $\theta_y$ [See Eq.~(\ref{eq})], leading to the conclusion that NM layers displaying large conventional spin Hall angles might prevent deterministic SOT switching in symmetry-broken bilayer heterostructures. Our analysis suggests that regime I type-of-dynamics is prevented in the limit $\tan(\theta_{\rm{tilt}}) \rightarrow \theta_z/\theta_y$ where 
	\begin{eqnarray}
		\frac{\theta_z}{\theta_y^2}  = \displaystyle \frac{\hbar}{4e}\beta\frac{J}{K_{\rm{eff}}d},
		\label{eq8}
	\end{eqnarray}
	beyond which a new dynamical regime, other than regimes I and II, must exist. We emphasize that Eq.~(\ref{eq8}) does not offer a quantitative description of the regime I boundary at larger $\theta_y$ values due to our initial assumptions of small $\theta_{\rm{tilt}}$. The generalization of Eq.~(\ref{eq8}) for arbitrary $\theta_y$ will be left for a future work. We note that this law described the $\beta > 0$ situation. As for the $\beta < 0$, the transition to regime II can be estimated by Eq.~(22) of Ref.~\cite{ref12}, which sets the limit $\theta_y$ for optimizing switching.
	
	Therefore, we conclude that the field-like torque due to the CSOT is the main agent behind the existence of stable precessional and steady states regimes. This observation is of fundamental importance in guiding the search of symmetry-broken spin Hall channels displaying both $\sigma_{zx}^y$ and $\sigma_{zx}^z$ for efficient switching of a perpendicular magnetization. Because field-like SOTs are always present in bilayer heterostructures, our results suggest that one should look into materials displaying large $\theta_z$ while maintaining $\theta_z/\theta_y$ limited to the regime I region of the state diagram in Fig.~\ref{fig4}(b). This condition necessarily implies the existence of an optimal $\theta_y$, contrary to the common belief of maximizing both $\theta_y$ and $\theta_z$ for efficient magnetization reversal. The deterministic switching optimization is discussed next.
	
	%We conclude that the field-like SOT are expected to substantially modify the transition between dynamical regimes I and II, with the possibility of intermediary dynamical states. In the following sections, we explore in detail transitions between regimes I and II by numerically solving the LLG equation. We assume the following parameters throughout: $\alpha = 0.001$, $d = 1$ nm, $K_{\rm{eff}} = 2 \times 10^5$ J/m$^3$ and $M_{\rm{S}} = 1.27\times 10^6$ A/m. 

	\section{VII. Optimizing deterministic Switching}
	
	Our previous results have established that full deterministic switching takes place within the limited region of the parameter space corresponding to regime I. We now address the question of how deterministic switching can be further optimized within this dynamical regime. To tackle this problem, we study the parameter dependence of the normalized energy delay product, $E_{\rm{sw}}\tau_{\rm{sw}} \propto (J\tau_{\rm{sw}})^2$, where $E_{\rm{sw}}$ and $\tau_{\rm{sw}}$ are the switching energy and time associated with a current density $J$. This figure of merit captures the application need for switching which is both low energy and fast. 
	
	\begin{figure}[t]
		\includegraphics[width =0.9\linewidth]{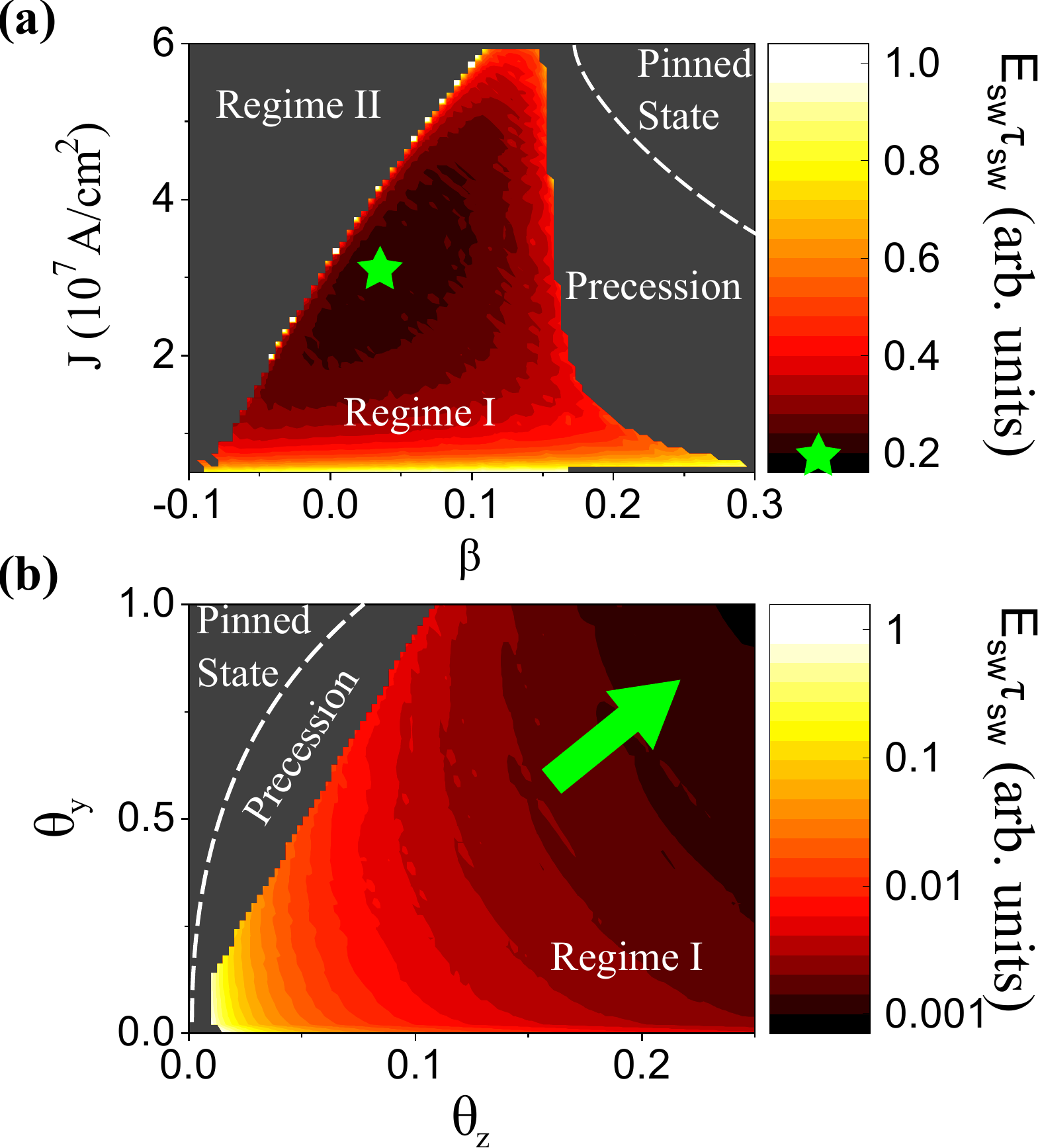}
		\caption{(Color online) Regime I normalized switching energy-delay product, $E_{\rm{sw}}\tau_{\rm{sw}}$, of a perpendicular magnetization subjected to both CSOT and USOT. (a) Current density vs field-like to damping-like torque ratio ($J$ vs $\beta$) diagram, at fixed $\theta_y = 0.6$ and $\theta_z = 0.05$. The energy-delay product was normalized by that of $J = 0.5 \times 10^7$ A/cm$^2$ and $\beta = 0.15$. The green star highlights the optimal energy-delay product. (b) $\theta_y$ vs $\theta_z$ diagram, at fixed $J = 3 \times 10^7$ A/cm$^2$ and $\beta = +0.2$, where the we normalized the energy-delay product by that of the case where $\theta_y = 0$ and $\theta_z = 0.015$. The green arrow highlights indicates the direction of best energy-delay product. The darker regions correspond to dynamical regimes other than regime I.}
		\label{fig5}
	\end{figure} 
	
	In the dynamical regime I, the switching time $\tau_{\rm{sw}}$ is obtained numerically by registering the time spent for full magnetization reversal according to the criteria $m_z(\tau_{\textrm{sw}}) < -0.95$ and $(dm_z/dt)_{t = \tau_{\rm{sw}}} < 0.01$, at every point of a given parameter space. The normalized energy-delay product in a $J$ vs $\beta$ state diagram, at fixed $\theta_y = 0.6$ and $\theta_z = 0.05$, is shown in Fig.~\ref{fig5}(a), where the darker regions correspond to dynamical regimes other than regime I. The result shows that there exists an optimum switching efficiency at positive $\beta$, where the energy-delay product can be reduced by approximately $50 \%$ of its maximum value [See darker blue region in Fig.~\ref{fig5}(a)]. Further, we notice that the associated optimum current density is higher than the regime I threshold switching current. This is due to the fact that the rate of decay of the switching time with the applied current density is faster than the rate of increase of $J$ at lower current density. However, the rate of decay of the switching time shows down near the regime I boundaries, such that the product $J \tau_{\rm{sw}}$ becomes an increasing function of $J$. Therefore, an optimum current density exists and corresponds to a minimum of $J\tau_{\rm{sw}}$.  
	
	Figure~\ref{fig5}(b) displays the normalized energy-delay product in a $\theta_y$ vs $\theta_z$ state diagram at fixed $J = 3 \times 10^7$ A/cm$^2$ and $\beta = +0.2$. The efficiency is normalized to that corresponding to the minimum spin Hall angles necessary to attain full switching at the given $J$ and $\beta$ values: $\theta_y = 0$ and $\theta_z = 0.015$. Here, efficiency is a monotonic function of the unconventional spin Hall angle with the possibility of attaining several orders of magnitude improvement upon increasing $\theta_z$. On the other hand, we also observe that virtually no efficiency improvement is attained at larger $\theta_y$. Instead, larger $\theta_y$ might actually prevent deterministic switching by inducing transitions to precessional states, as previously discussed. 
	
	The above results establish that the largest source of efficiency improvement within regime I is the unconventional spin Hall angle. Although an optimized current density can be found, deterministic switching becomes much more efficient in bilayers composed of materials displaying larger $\theta_z$ values and an optimal ratio $\theta_z/\theta_y$. A few tentative bilayer structures composed of symmetry-broken NMs potentially suitable for observing the effects mentioned above are CoFeB(Py)/BiTe$_3$, CoFeB(Py)/Ni$_2$P$_6$W$_4$, CoFeB(Py)/Ba$_2$C$_4$S$_4$N$_4$~\cite{SHEDataBase} or Py/SrIrO$_3$. The latter was recently investigated experimentally in the context of CSOTs~\cite{SrIrO1} with the prediction of giant unconventional spin Hall effects in orthorhombic SrIrO$_3$~\cite{SrIrO2}.
	
	\section{VIII. Conclusions}
	
	We have studied the simultaneous impact of CSOT and USOT on the dynamics of a perpendicular magnetization in bilayer heterostructures in the small intrinsic damping limit. Our results reveal that the field-like CSOT immensely impacts the dynamical state of the perpendicular magnetization, leading to the formation of distinctive dynamical regimes over relevant parameter spaces. Particularly, we showed the existence of unexpected stable precessional and steady states at larger conventional spin Hall angles. This implies that non-magnets displaying large conventional spin Hall angles, $\theta_y$, might be detrimental to full deterministic switching in this system. We also establish that highly efficient deterministic switching is attained at larger unconventional spin Hall angles, $\theta_z$, within a specific dynamical regime constrained by the ratio $\theta_z/ \theta_y$. Our work clarifies the physics and performance metric necessary for highly energy-efficient switching of a perpendicular magnetization in symmetry broken bilayer heterostructures.
	%Our work offers a systematic path to realize highly energy efficient SOT switching in symmetry broken bilayer heterostructures.

	\section{Acknowledgments}D.S, J.P.W and T.L were partially supported by the SMART, one of seven centers of nCORE, a Semiconductor Research Corporation program, sponsored by National Institute of Standards and Technology (NIST) and by the DARPA ERI FRANC program under HR001117S0056-FP-042. D.S thanks T. J. Peterson for helpful discussions.


\begin{thebibliography}{apsrev}
\bibitem{footnote}
		We emphasize that the sign convention for $\beta$ is tightly connected to the cross product orderings of Eq.~(\ref{eq1}), which have not been consistent among different authors~\cite{ref11, ref12}.
\bibitem{footnote2}
		This simplification is enabled by the fact that at sufficiently small $\theta_{\rm{tilt}}$, the overall damping profile over the unit sphere due to SOT and USOT is approximately the same as that of the case where $\theta_{\rm{tilt}} = 0$ up to a small rotation of the coordinate system around the $x$ axis.
\end{thebibliography}
\end{document}